\documentstyle[12pt,epsf]{article}

\newcommand{\agt}{\,\rlap{\lower 3.5 pt \hbox{$\mathchar \sim$}} \raise 1pt
 \hbox {$>$}\,}
\newcommand{\alt}{\,\rlap{\lower 3.5 pt \hbox{$\mathchar \sim$}} \raise 1pt
 \hbox {$<$}\,}

\begin{document}

\title{\vskip-3cm{\baselineskip14pt
\centerline{\normalsize\hfill TTP98--21}
\centerline{\normalsize\hfill NYU--TH/98/04/02}
\centerline{\normalsize\hfill MPI/PhT/98--032}
\centerline{\normalsize\hfill FERMILAB--PUB--98/126--T}
\centerline{\normalsize\hfill hep--ph/9807241}
\centerline{\normalsize\hfill June 1998}
}
\vskip1.5cm
Effective QCD Interactions of CP-odd Higgs Bosons at Three Loops
}
\author{K.G. Chetyrkin$^{1,}$\thanks{Permanent address:
Institute for Nuclear Research, Russian Academy of Sciences,
60th October Anniversary Prospect 7a, Moscow 117312, Russia.},
B.A. Kniehl$^{2,}$\thanks{Permanent address:
Max-Planck-Institut f\"ur Physik (Werner-Heisenberg-Institut),
F\"ohringer Ring~6, 80805 Munich, Germany.},
M. Steinhauser$^3$, and W.A. Bardeen$^4$\\
{\normalsize $^1$
Institut f\"ur Theoretische Teilchenphysik, Universit\"at Karlsruhe,}\\
{\normalsize Kaiserstra\ss e 12, 76128 Karlsruhe, Germany}\\
{\normalsize $^2$
Department of Physics, New York University,}\\
{\normalsize 4 Washington Place, New York, NY 10003, USA}\\
{\normalsize $^3$
Max-Planck-Institut f\"ur Physik (Werner-Heisenberg-Institut),}\\
{\normalsize F\"ohringer Ring 6, 80805 Munich, Germany}\\
{\normalsize $^4$
Theoretical Physics Department, Fermi National Accelerator Laboratory,}\\
{\normalsize P.O. Box 500, Batavia, IL 60510, USA}
}

\date{}

\maketitle

\thispagestyle{empty}

\begin{abstract}
In the virtual presence of a heavy quark $t$, the interactions of a CP-odd 
scalar boson $A$, with mass $M_A\ll2M_t$, with gluons and light quarks can be
described by an effective Lagrangian.
We analytically derive the coefficient functions of the respective physical
operators to three loops in quantum chromodynamics (QCD), adopting the
modified minimal-subtraction ($\overline{\rm MS}$) scheme of dimensional 
regularization.
Special attention is paid to the proper treatment of the $\gamma_5$ matrix and 
the Levi-Civita $\epsilon$ tensor in $D$ dimensions.
In the case of the effective $ggA$ coupling, we find agreement with an
all-order prediction based on a low-energy theorem in connection with the
Adler-Bardeen non-renormalization theorem.
This effective Lagrangian allows us to analytically evaluate the
next-to-leading QCD correction to the $A\to gg$ partial decay width by
considering massless diagrams.
For $M_A=100$~GeV, the resulting correction factor reads
$1+(221/12)\alpha_s^{(5)}(M_A)/\pi
+165.9\left(\alpha_s^{(5)}(M_A)/\pi\right)^2\approx1+0.68+0.23$.
We compare this result with predictions based on various scale-optimization
methods.
\medskip

\noindent
PACS numbers: 11.15.Me, 12.38.Bx, 13.30.Eg, 14.80.Cp
\end{abstract}

\newpage

\section{Introduction}

Despite the tremendously successful consolidation of the standard model (SM)
of elementary particle physics by experimental precision tests during the past
few years, the structure of the Higgs sector has essentially remained
unexplored, so that there is still plenty of room for extensions.
A phenomenologically interesting extension of the SM Higgs sector that keeps
the electroweak $\rho$ parameter \cite{vel} at unity in the Born
approximation, is obtained by adding a second complex isospin-doublet scalar
field with opposite hypercharge.
This leads to the two-Higgs-doublet model (2HDM).
After the three massless Goldstone bosons which emerge via the electroweak
symmetry breaking are eaten up to become the longitudinal degrees of freedom
of the $W^\pm$ and $Z$ bosons, there remain five physical Higgs scalars: the
neutral CP-even $h$ and $H$ bosons, the neutral CP-odd $A$ boson, and the
charged $H^\pm$-boson pair.
The Higgs sector of the minimal supersymmetric extension of the SM (MSSM)
consists of such a 2HDM.
At tree level, the MSSM Higgs sector has two free parameters, which are
usually taken to be the mass $M_A$ of the $A$ boson and the ratio
$\tan\beta=v_2/v_1$ of the vacuum expectation values of the two Higgs 
doublets.
For large values of $\tan\beta$, the top Yukawa couplings of the neutral Higgs 
bosons, $\Phi=h,H,A$, are suppressed compared to the bottom ones.

The search for Higgs bosons and the study of their properties are among the 
prime objectives of the Large Hadron Collider (LHC), a proton-proton
colliding-beam facility with centre-of-mass energy $\sqrt s=14$~GeV, which is
presently under construction at CERN.
The dominant production mechanisms for the neutral Higgs bosons at the LHC
will be gluon fusion, $gg\to\Phi$ \cite{geo}, and $b\bar b\Phi$ associated
production, $gg,q\bar q\to b\bar b\Phi$ \cite{bar}, which is, however, only
relevant for large $\tan\beta$.
The loop-induced $gg\Phi$ couplings \cite{wil} are mainly mediated by virtual
top quarks, unless $\tan\beta$ is very large, in which case the bottom-quark
loops take over.
The $ggh$ and $ggH$ couplings also receive contributions from squark loops,
which are, however, insignificant for squark masses in excess of about 500~GeV
\cite{daw}.
In the case of the $ggA$ coupling, such contributions do not occur at one loop
because the $A$ boson has no tree-level couplings to squarks.
For small $\tan\beta$, the inclusive cross sections of $pp\to\Phi+X$ via
gluon fusion are significantly increased, by typically 50--70\% under LHC
conditions, by including their leading QCD corrections, which involve two-loop
contributions \cite{sda,spi}.
Thus, the theoretical predictions for these observables cannot yet be
considered to be well under control, and it is desirable to compute the
next-to-leading QCD corrections at three loops, since there is no reason to
expect them to be negligible.
Recently, a first step in this direction has been taken by considering the
resummation of soft-gluon radiation in $pp\to\Phi+X$, assuming $\tan\beta$ to
be small \cite{kra}.

Considering the enormous complexity of the exact expressions for the leading
QCD corrections \cite{spi}, it becomes apparent that, with presently available
technology, the next-to-leading corrections are only tractable in limiting
cases.
For instance, in the large-$\tan\beta$ limit, where the $gg\Phi$ couplings
are chiefly generated by bottom-quark loops, one can neglect the bottom-quark
mass against the Higgs-boson mass, keeping the bottom Yukawa coupling finite.
In this way, one resorts to massless QCD.
On the other hand, if $\tan\beta$ is close to unity, which was assumed in 
Ref.~\cite{kra}, the top-quark loops play the dominant r\^ole, and 
simplifications occur if the mass hierarchy $M_\Phi\ll2m_t$ is satisfied.
Then, it is useful to construct an effective Lagrangian by integrating out the
top quark.
This effective Lagrangian is a linear combination of local composite operators
of mass dimension four, which act in QCD with five massless quark flavours,
while all dependence on $m_t$ is contained in their coefficient functions.
Once the coefficient functions are known, it is sufficient to deal with
massless Feynman diagrams.
The effective Lagrangian describing the interactions of the SM Higgs boson $H$
with gluons and light quarks was elaborated at two loops in Ref.~\cite{ina}
and extended to three loops in Refs.~\cite{che,dec}.
As an application, the ${\cal O}(\alpha_s)$ \cite{ina} and
${\cal O}(\alpha_s^2)$ \cite{che} corrections to the $H\to gg$ partial decay
width were calculated from this Lagrangian.
These results can be readily adapted to the $h$ and $H$ bosons of the 2HDM
with small $\tan\beta$ by accordingly adjusting the top Yukawa coupling, which
appears as an overall factor.

In this paper, we extend the three-loop analysis of Refs.~\cite{che,dec} to
include the $A$ boson of the 2HDM.
As in Ref.~\cite{kra}, we work in the limit $\tan\beta\approx1$, so that we
may treat bottom as a massless quark flavour with vanishing Yukawa coupling,
on the same footing as up, down, strange, and charm.
Specifically, we construct a heavy-top-quark effective Lagrangian for the QCD
interactions of the $A$ boson and derive from it an analytic result for the
${\cal O}(\alpha_s^2)$ correction to the $A\to gg$ partial decay width
appropriate for $M_A\ll2m_t$.
We recover the corresponding ${\cal O}(\alpha_s)$ result originally found in
Refs.~\cite{spi,kat} and also discussed in Ref.~\cite{zer}.
The ${\cal O}(\alpha_s)$ correction for arbitrary values of the $A$-boson and
quark masses was presented in Ref.~\cite{spi} as a two-fold parameter
integral.
Furthermore, it was shown that the leading high-$m_t$ term of this correction
may also be obtained from massless five-flavour QCD endowed with a
heavy-top-quark effective $ggA$ coupling \cite{spi,djo}.
A central ingredient for this check was the observation that the effective
$ggA$ coupling does not receive QCD corrections, at least at
${\cal O}(\alpha_s)$.
This fact was interpreted \cite{spi,djo} as being a consequence of the
Adler-Bardeen theorem \cite{adl}, which states that the anomaly of the
axial-vector current \cite{abj} is not renormalized in QCD.
This theorem is strictly proven to all orders in $\alpha_s$ for the abelian
case \cite{adl}, and strong arguments suggest that it also holds true for the
nonabelian case \cite{wab}.
In this paper, we verify by an explicit diagrammatic calculation that the
${\cal O}(\alpha_s)$ and ${\cal O}(\alpha_s^2)$ corrections to the coefficient
function of the operator $G_{\mu\nu}^a\tilde G^{a\mu\nu}$, which generates the
$A$-boson effective couplings to gluons vanish.\footnote{Strictly speaking,
this statement is only true as long as we ignore the axial-anomaly equation,
as will become apparent in Section~\ref{sec:four}.}
We also present the leading-order coefficient function of the physical
operator pertaining to the effective $q\bar qA$ interaction, where $q$ is a
light quark.
We thus provide the tools which are necessary to reduce the calculation of the
next-to-leading QCD correction to the cross section of $pp\to A+X$ to a
standard problem in massless five-flavour QCD.

In our analysis, we consistently neglect the Yukawa couplings of the light 
quarks to the $A$ boson.
In other words, if all quark masses, except for $m_t$, are nullified, the
hadronic decay width of the $A$ boson is entirely due to $A\to gg$ and the
associated higher-order processes under consideration here.
Through three loops, the contributing final states are $gggg$, $ggq\bar q$,
$q\bar qq^\prime\bar q^\prime$, $ggg$, $gq\bar q$, $gg$, and $q\bar q$.

This paper is organized as follows.
In Section~\ref{sec:two}, we establish the heavy-top-quark effective
Lagrangian for the QCD interactions of the $A$ boson to three loops.
In Section~\ref{sec:three}, we compute from this Lagrangian the
${\cal O}(\alpha_s^2)$ correction to the partial width of the decay
$A\to gg$ and compare it with predictions based on various
scale-optimization methods.
Section~\ref{sec:four} contains a discussion of our results together with
some remarks on the connection between the effective $ggA$ coupling, the
axial-anomaly equation, and the low-energy theorem.

\section{\label{sec:two}Effective Lagrangian}

We start by setting up the theoretical framework for our analysis.
As usual, we employ dimensional regularization in $D=4-2\varepsilon$
space-time dimensions and introduce a 't~Hooft mass, $\mu$, to keep the
coupling constants dimensionless \cite{bol,tho}.
We perform the renormalization according to the modified \cite{bur}
minimal-subtraction \cite{hoo} ($\overline{\rm MS}$) scheme.
For the sake of generality, we take the QCD gauge group to be SU($N_c$), with
$N_c$ arbitrary.
The adjoint representation has dimension $N_A=N_c^2-1$.
The colour factors corresponding to the Casimir operators of the fundamental
and adjoint representations are $C_F=(N_c^2-1)/(2N_c)$ and $C_A=N_c$,
respectively.
For the numerical evaluation, we set $N_c=3$.
The trace normalization of the fundamental representation is $T=1/2$.
As an idealized situation, we consider QCD with $n_l=n_f-1$ light quark
flavours $q_i$ and one heavy flavour $t$, in the sense that
$2m_{q_i}\ll M_A\ll 2m_t$.
We wish to construct an effective $n_l$-flavour theory by integrating out the
$t$ quark.
We mark the quantities of the effective theory by a prime.
Bare quantities carry the superscript ``0".
As already mentioned in the Introduction, we consider a 2HDM with 
$\tan\beta=1$, so that the quark Yukawa couplings and masses are related by a
flavour-independent proportionality factor.

The starting point of our consideration is the bare Yukawa Lagrangian for the 
interactions of the $A$ boson with the quarks in the full $n_f$-flavour
theory,
\begin{equation}
{\cal L}=-\frac{A^0}{v^0}\left(\sum_{i=1}^{n_l}
m_{q_i}^0\bar q_i^0i\gamma_5q_i^0+m_t^0\bar t^0i\gamma_5t^0\right),
\label{ful}
\end{equation}
where $v=2^{-1/4}G_F^{-1/2}$, with $G_F$ being Fermi's constant.
Taking the limit $m_t^0\to\infty$ and keeping only leading terms, 
Eq.~(\ref{ful}) may be written as a linear combination of pseudoscalar
composite operators, $\tilde O_i^\prime$, with mass dimension four acting in
the effective $n_l$-flavour theory.
The resulting bare Lagrangian reads
\begin{equation}
{\cal L}_{\rm eff}=-\frac{A^0}{v^0}\left(
\tilde C_1^0\tilde O_1^\prime+\tilde C_2^0\tilde O_2^\prime+\ldots\right),
\label{eff}
\end{equation}
where
\begin{eqnarray}
\tilde O_1^\prime&=&G_{\mu\nu}^{0\prime,a}\tilde G^{0\prime,a\mu\nu},
\nonumber\\
\tilde O_2^\prime&=&\partial_\mu J_5^{0\prime,\mu},\qquad
J_5^{0\prime,\mu}=\sum_{i=1}^{n_l}\bar q_i^{0\prime}\gamma^\mu\gamma_5
q_i^{0\prime},
\label{ope}
\end{eqnarray}
$\tilde C_i^0$ are coefficient functions, which depend on the bare parameters
of the full theory and carry all $m_t^0$ dependence, and the ellipsis stands
for terms involving unphysical operators, which do not contribute to physical
observables.
Here, $G_{\mu\nu}^a=\partial_\mu G_\nu^a-\partial_\nu G_\mu^a
+g_sf^{abc}G_\mu^bG_\nu^c$ is the colour-field-strength tensor and
$\tilde G^{a\mu\nu}=\epsilon^{\mu\nu\rho\sigma}G_{\rho\sigma}^a$ is its dual;
$G_\mu^a$ ($a=1,\ldots,N_A$) are the gluon fields, $g_s=\sqrt{4\pi\alpha_s}$
is the QCD gauge coupling, and $f^{abc}$ are the structure constants of the
SU($N_c$) algebra.
We do not display the colour indices of the quark fields.
We mark the operators and coefficient functions with a tilde in order to avoid
confusion with our previous notation for the scalar case \cite{che,dec}.

The Levi-Civita tensor $\epsilon^{\mu\nu\rho\sigma}$ is unavoidably a 
four-dimensional object and should be taken outside the R operation.
Thus, we rewrite Eq.~(\ref{ope}) as
$\tilde O_i^\prime=\epsilon^{\mu\nu\rho\sigma}
\tilde O_{i,\mu\nu\rho\sigma}^\prime$, where \cite{tho,aky,lar}
\begin{eqnarray}
\tilde O_{1,\mu\nu\rho\sigma}^\prime&=&
G_{[\mu\nu}^{0\prime,a}G_{\rho\sigma]}^{0\prime,a},\nonumber\\
\tilde O_{2,\mu\nu\rho\sigma}^\prime&=&
\frac{i}{3!}\sum_{i=1}^{n_l}\partial_{[\mu}\bar q_i^{0\prime}
\gamma_\nu\gamma_\rho\gamma_{\sigma]}q_i^{0\prime}
\label{lev}
\end{eqnarray}
are antisymmetrized in their four $D$-dimensional Lorentz indices.
Furthermore, we substitute $\gamma_5=(i/4!)\epsilon^{\mu\nu\rho\sigma}
\gamma_{[\mu}\gamma_\nu\gamma_\rho\gamma_{\sigma]}$ in Eq.~(\ref{ful}).
We then carry out the $D$-dimensional calculations with 
$\epsilon^{\mu\nu\rho\sigma}$ peeled off from the expressions.
In the very end, after the renormalization is performed and the physical limit
$\varepsilon\to0$ is taken, we contract the expressions with 
$\epsilon^{\mu\nu\rho\sigma}$ to obtain the final results.

Prior to describing the actual calculation of $\tilde C_i^0$, let us discuss
how Eq.~(\ref{eff}) is renormalized.
Since we are only interested in pure QCD corrections, we may substitute
$A^0=A$ and $v^0=v$ in Eqs.~(\ref{ful}) and (\ref{eff}).
Denoting the renormalized counterparts of $\tilde C_i^0$ and
$\tilde O_i^\prime$ by $\tilde C_i$ and $[\tilde O_i^\prime]$, the
renormalized version of Eq.~(\ref{eff}) takes the form
\begin{equation}
{\cal L}_{\rm eff}=-2^{1/4}G_F^{1/2}A\left(
\tilde C_1\left[\tilde O_1^\prime\right]
+\tilde C_2\left[\tilde O_2^\prime\right]+\ldots\right),
\label{ren}
\end{equation}
where the ellipsis again represents unphysical terms.
The divergence $\partial_\mu J_5^{0\prime,\mu}$ is renormalized
multiplicatively in the same way as the colour-singlet axial-vector current
$J_5^{0\prime,\mu}$ itself, while
$G_{\mu\nu}^{0\prime,a}\tilde G^{0\prime,a\mu\nu}$ mixes under renormalization
\cite{lar,esp}.
Specifically, we have 
\begin{eqnarray}
\left[\tilde O_1^\prime\right]&=&
Z_{11}^\prime\tilde O_1^\prime+Z_{12}^\prime\tilde O_2^\prime,
\nonumber\\
\left[\tilde O_2^\prime\right]&=&Z_{22}^\prime\tilde O_2^\prime,
\label{mix}
\end{eqnarray}
where $Z_{22}^\prime=Z_{\rm MS}^{\rm s\prime}Z_5^{\rm s\prime}$ is the product
of the standard ultraviolet (UV) renormalization constant
$Z_{\rm MS}^{\rm s\prime}$ of the singlet axial current in the
$\overline{\rm MS}$ scheme and the finite renormalization constant
$Z_5^{\rm s\prime}$.
The latter is introduced to restore the one-loop character of the operator
relation of the axial anomaly,
\begin{equation}
\left[\tilde O_2^\prime\right]=
\frac{\alpha_s^{(n_l)}(\mu)}{\pi}\,\frac{Tn_l}{4}
\left[\tilde O_1^\prime\right],
\label{ano}
\end{equation}
which is valid for Pauli-Villars regularization \cite{lar}.
For the reader's convenience, we list here the various renormalization
constants to the order necessary for our purposes.
They read \cite{lar}
\begin{eqnarray}
Z_{11}^\prime&=&
1+\frac{\alpha_s^{(n_l)}(\mu)}{\pi}\,\frac{1}{\varepsilon}
\left(-\frac{11}{12}C_A+\frac{1}{3}Tn_l\right)
+\left(\frac{\alpha_s^{(n_l)}(\mu)}{\pi}\right)^2
\nonumber\\
&&{}\times
\left[\frac{1}{\varepsilon^2}
\left(\frac{121}{144}C_A^2-\frac{11}{18}C_ATn_l+\frac{1}{9}T^2n_l^2\right)
+\frac{1}{\varepsilon}
\left(-\frac{17}{48}C_A^2+\frac{1}{8}C_FTn_l+\frac{5}{24}C_ATn_l\right)\right],
\nonumber\\
Z_{12}^\prime&=&\frac{\alpha_s^{(n_l)}(\mu)}{\pi}\,\frac{1}{\varepsilon}
(3C_F),
\nonumber\\
Z_{\rm MS}^{\rm s\prime}&=&Z_5^{\rm s\prime}=1.
\end{eqnarray}
Note that $Z_{11}^\prime=\left(Z_g^\prime\right)^2$ is the square of the
coupling renormalization constant, $Z_g^\prime$.
As will become apparent later, we only need the leading term of
$Z_{22}^\prime$.
The relations between the bare and the renormalized coefficient functions are
accordingly given by
\begin{eqnarray}
\tilde C_1&=&\frac{1}{Z_{11}^\prime}\tilde C_1^0,\nonumber\\
\tilde C_2&=&\frac{1}{Z_{22}^\prime}
\left(-\frac{Z_{12}^\prime}{Z_{11}^\prime}\tilde C_1^0+\tilde C_2^0\right).
\label{coe}
\end{eqnarray}

\begin{figure}[t]
\leavevmode
\begin{center}
\epsfxsize=16cm
\epsffile[76 635 552 724]{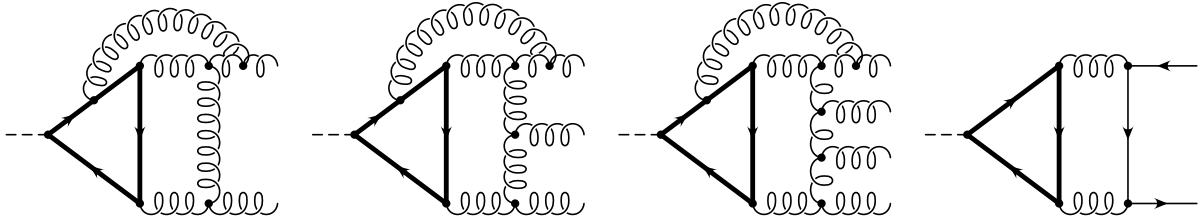}
\caption{Typical Feynman diagrams contributing to the coefficients
$\tilde C_i^0$ in Eq.~(\ref{eff}).
Looped, bold-faced, and dashed lines represent gluons, $t$ quarks, and $A$ 
bosons, respectively.}
\label{fig:one}
\end{center}
\end{figure}

We now turn to the computation of the bare coefficient functions
$\tilde C_1^0$ and $\tilde C_2^0$ in Eq.~(\ref{eff}).
We are thus led to consider irreducible vertex diagrams which connect one $A$
boson to the respective states of gluons and light-quark pairs via one or more
$t$-quark loops, whereby all external particles are taken to be on their mass
shells.
Typical examples are depicted in Fig.~\ref{fig:one}.
There are three independent ways to obtain $\tilde C_1^0$, namely from the
sets of $ggA$ three-point, $gggA$ four-point, or $ggggA$ five-point diagrams.
At the three-loop level, these sets contain 657, 7362, and 95004 diagrams, 
respectively.
We choose to work out the first option in the covariant gauge with arbitrary
gauge parameter, so that the gauge-parameter independence of the final result
yields a nontrivial check.
Another independent check is then provided by elaborating the second option in 
the 't~Hooft-Feynman gauge keeping only one external momentum different from 
zero.
In order to cope with the enormous complexity of the problem at hand, we make
successive use of powerful symbolic manipulation programs.
Specifically, we generate and evaluate the contributing diagrams with the
packages QGRAF \cite{nog} and MATAD \cite{ste}, which is written in FORM
\cite{ver}, respectively.
The cancellation of the UV singularities, the gauge-parameter independence,
and the renormalization-group (RG) invariance serve as strong checks for our
calculation.

Let us denote the sum of all relevant $ggA$ diagrams by
$V_{ggA,\alpha\beta}^{\mu\nu\rho\sigma}(q_1,q_2)$, where $q_1$ and $q_2$ are
the incoming four-momenta of the two on-shell gluons with polarization
four-vectors $\epsilon_1^\alpha$ and $\epsilon_2^\beta$, respectively.
According to Eq.~(\ref{lev}), $V_{ggA,\alpha\beta}^{\mu\nu\rho\sigma}$ is by
construction totally antisymmetric in the indices $\mu$, $\nu$, $\rho$, and
$\sigma$.
In order to compute $\tilde C_1^0$, we need to expand $V_{ggA}$ up to terms
linear in $q_1$ and $q_2$.
There is just one possible structure, namely
\begin{equation}
V_{ggA,\alpha\beta}^{\mu\nu\rho\sigma}(q_1,q_2)
=C_{ggA}P_{ggA,\alpha\beta}^{\mu\nu\rho\sigma}(q_1,q_2),\qquad
P_{ggA,\alpha\beta}^{\mu\nu\rho\sigma}(q_1,q_2)=
q_1^{[\mu}q_2^\nu g_\alpha^\rho g_\beta^{\sigma]},
\end{equation}
so that the coefficient $C_{ggA}$ may be conveniently extracted by noting that
$P_{ggA}^2=-(D-2)(D-3)(q_1\cdot q_2)^2/24$.
The final formula for $\tilde C_1^0$ reads
\begin{equation}
\tilde C_1^0=-\frac{3}{(D-2)(D-3)(q_1\cdot q_2)^2}\,
\frac{Z_{\rm MS}^{\rm p}Z_5^{\rm p}}{Z_m\zeta_3^0}
P_{ggA,\mu\nu\rho\sigma}^{\alpha\beta}(q_1,q_2)
V_{ggA,\alpha\beta}^{\mu\nu\rho\sigma}(q_1,q_2).
\label{one}
\end{equation}
The normalization of Eq.~(\ref{one}) may be understood as follows.
On the one hand, we need to renormalize the mass and the pseudoscalar current 
of the $t$ quark in the Lagrangian~(\ref{ful}) of the $n_f$-flavour theory.
In this way, the renormalization constants $Z_m$ \cite{tar},
$Z_{\rm MS}^{\rm p}$, and $Z_5^{\rm p}$ \cite{lar} enter.
They are defined as
\begin{eqnarray}
m_t^0&=&Z_mm_t,\nonumber\\
\left[\bar t\gamma_5t\right]&=&
Z_{\rm MS}^{\rm p}Z_5^{\rm p}\bar t^0\gamma_5t^0,
\end{eqnarray}
and read \cite{lar,tar}
\begin{eqnarray}
Z_m&=&1+\frac{\alpha_s^{(n_f)}(\mu)}{\pi}\,
\frac{1}{\varepsilon}\left(-\frac{3}{4}C_F\right)
+\left(\frac{\alpha_s^{(n_f)}(\mu)}{\pi}\right)^2
\left[\frac{1}{\varepsilon^2}
\left(\frac{9}{32}C_F^2+\frac{11}{32}C_FC_A-\frac{1}{8}C_FTn_f\right)
\right.\nonumber\\
&&{}+\left.
\frac{1}{\varepsilon}\left(-\frac{3}{64}C_F^2-\frac{97}{192}C_FC_A
+\frac{5}{48}C_FTn_f\right)\right],
\nonumber\\
Z_{\rm MS}^{\rm p}&=&1+\frac{\alpha_s^{(n_f)}(\mu)}{\pi}\,
\frac{1}{\varepsilon}\left(-\frac{3}{4}C_F\right)
+\left(\frac{\alpha_s^{(n_f)}(\mu)}{\pi}\right)^2
\left[\frac{1}{\varepsilon^2}
\left(\frac{9}{32}C_F^2+\frac{11}{32}C_FC_A-\frac{1}{8}C_FTn_f\right)
\right.\nonumber\\
&&{}+\left.
\frac{1}{\varepsilon}\left(-\frac{3}{64}C_F^2+\frac{79}{192}C_FC_A
-\frac{11}{48}C_FTn_f\right)\right],
\nonumber\\
Z_5^{\rm p}&=&1+\frac{\alpha_s^{(n_f)}(\mu)}{\pi}(-2C_F)
+\left(\frac{\alpha_s^{(n_f)}(\mu)}{\pi}\right)^2
\left(\frac{1}{72}C_FC_A+\frac{1}{18}C_FTn_f\right).
\end{eqnarray}
The finite renormalization constant $Z_5^{\rm p}$ is needed in addition to the
usual UV renormalization constant of the $\overline{\rm MS}$ scheme,
$Z_{\rm MS}^{\rm p}$, to effectively restore the anticommutativity of the 
$\gamma_5$ matrix \cite{lar,tru}.
On the other hand, we need to express the bare couplings and fields appearing
in the Lagrangian~(\ref{eff}) of the $n_l$-flavour theory in terms of their
counterparts in the $n_f$-flavour theory.
The appropriate relations,
\begin{equation}
g_s^{0\prime}=\zeta_g^0g_s^0,\qquad
q_i^{0\prime}=\sqrt{\zeta_2^0}q_i^0,\qquad
G_\mu^{0\prime,a}=\sqrt{\zeta_3^0}G_\mu^{0,a},
\end{equation}
involve the decoupling constants $\zeta_g^0$, $\zeta_2^0$, and $\zeta_3^0$.
In the case of the $ggA$ amplitude, only $\zeta_3^0$ occurs.
For our purposes, we need $\zeta_3^0$ through
${\cal O}(\varepsilon^2\alpha_s)$ and ${\cal O}(\varepsilon\alpha_s^2)$.
The corresponding expression may be extracted from Eq.~(B.2) of
Ref.~\cite{dec}, where the renormalized version, $\zeta_3$, is listed through
${\cal O}(\alpha_s^3)$ in the covariant gauge.
Finally, a factor of 8 stems from the Feynman rule for the two-gluon piece of
$\tilde O_1^\prime$.
In Eq.~(\ref{one}), we may take the limit $q_1,q_2\to0$, which reduces the
problem of finding $\tilde C_1^0$ to the solution of massive vacuum integrals
\cite{sgg}.
After renormalization according to Eq.~(\ref{coe}), we find
\begin{equation}
\tilde C_1=-\frac{\alpha_s^{(n_l)}(\mu)}{\pi}\,\frac{T}{8}
\left[1+0\cdot\frac{\alpha_s^{(n_l)}(\mu)}{\pi}
+0\cdot\left(\frac{\alpha_s^{(n_l)}(\mu)}{\pi}\right)^2\right],
\label{con}
\end{equation}
i.e.\ the correction terms of ${\cal O}(\alpha_s)$ and ${\cal O}(\alpha_s^2)$
indeed vanish, as was suggested in Ref.~\cite{spi,djo} on the basis of the
nonabelian variant \cite{wab} of the Adler-Bardeen theorem \cite{adl}, which
predicts that $\tilde C_1$ does not receive QCD corrections.
Notice that this is only true if $\tilde C_1$ is expanded in
$\alpha_s^{(n_l)}(\mu)$.

As an independent check, we may also extract $\tilde C_1^0$ from the sum
$V_{gggA,\alpha\beta\gamma}^{\mu\nu\rho\sigma}(q)$ of the $gggA$ diagrams,
where $\alpha$, $\beta$, and $\gamma$ are the Lorentz indices of the gluon
polarization four-vectors.
Notice that it is sufficient to keep one external four-momentum, $q$, 
different from zero, since the three-gluon piece of $\tilde O_1^\prime$ 
contains just one derivative.
Again, there is only one possible structure linear in $q$, namely
\begin{equation}
V_{gggA,\alpha\beta\gamma}^{\mu\nu\rho\sigma}(q)
=C_{gggA}P_{gggA,\alpha\beta\gamma}^{\mu\nu\rho\sigma}(q),\qquad
P_{gggA,\alpha\beta\gamma}^{\mu\nu\rho\sigma}(q)=
q^{[\mu}g_\alpha^\nu g_\beta^\rho g_\gamma^{\sigma]},
\end{equation}
so that the coefficient $C_{gggA}$ may be easily extracted by using
$P_{gggA}^2=(D-1)(D-2)(D-3)q^2/24$.
In contrast to the two-gluon piece of $\tilde O_1^\prime$, we now have to 
include decoupling constants for three gluon fields and one gauge coupling.
Again, the Feynman rule for the three-gluon piece of $\tilde O_1^\prime$
involves a factor of 8.
Thus, the final formula for $\tilde C_1^0$ is given by
\begin{equation}
\tilde C_1^0=\frac{3}{(D-1)(D-2)(D-3)q^2}\,
\frac{Z_{\rm MS}^{\rm p}Z_5^{\rm p}}{Z_m\zeta_g^0(\zeta_3^0)^{3/2}}
P_{gggA,\mu\nu\rho\sigma}^{\alpha\beta\gamma}(q)
V_{gggA,\alpha\beta\gamma}^{\mu\nu\rho\sigma}(q).
\label{alt}
\end{equation}
Taking the limit $q\to0$ and and working in the 't~Hooft-Feynman gauge, we
recover from Eq.~(\ref{alt}) our previous result for $\tilde C_1^0$.

Finally, we turn to $\tilde C_2^0$, which is generated by vertex diagrams that
couple the $A$ boson to a pair of light quarks and involve a virtual $t$
quark.
This requires at least two loops.
Specifically, there are 2 two-loop and 63 three-loop diagrams of this type.
Calling the resulting $q_i\bar q_iA$ amplitude
$V_{qqA}^{\mu\nu\rho\sigma}(q)$, where the argument $q$ is the incoming
four-momentum of the $A$ boson, we may extract $\tilde C_2^0$ as
\begin{equation}
\tilde C_2^0=\frac{6}{(D-1)(D-2)(D-3)q^2}\,
\frac{Z_{\rm MS}^{\rm p}Z_5^{\rm p}}{Z_m\zeta_2^0}
{\rm Tr}\,\left[q_\mu\gamma_\nu\gamma_\rho\gamma_\sigma 
V_{qqA}^{\mu\nu\rho\sigma}(q)\right].
\end{equation}

In order to treat the $A\to gg$ decay at three loops, it is sufficient to know
the leading term of $\tilde C_2$.
Moreover, due to Eq.~(\ref{coe}), the computation of the next-to-leading term
of $\tilde C_2$ would require the knowledge of $Z_{12}^\prime$ to
${\cal O}(\alpha_s^2)$, which is not yet available.
After renormalization according to Eq.~(\ref{coe}), we find
\begin{equation}
\tilde C_2=\left(\frac{\alpha_s^{(n_l)}(\mu)}{\pi}\right)^2C_FT
\left(\frac{3}{16}-\frac{3}{8}\ln\frac{\mu^2}{m_t^2(\mu)}\right),
\label{ctw}
\end{equation}
where $m_t(\mu)$ is the $\overline{\rm MS}$ mass of the $t$ quark.

\boldmath
\section{\label{sec:three}$A\to gg$ decay}
\unboldmath

Having established the high-$m_t$ effective Lagrangian~(\ref{ren}) controlling
the QCD interactions of the $A$ boson, we are now in a position to evaluate
from it the ${\cal O}(\alpha_s^2)$ correction to the $A\to gg$ decay width.
To this end, we need to compute the absorptive part of the $A$-boson
self-energy, at $q^2=M_A^2$, induced by $[\tilde O_1^\prime]$ and
$[\tilde O_2^\prime]$ to sufficiently high order in the $n_l$-flavour theory.

In $D$ dimensions, the correlator function, at four-momentum $q$, of two bare
operators of the type defined in Eq.~(\ref{lev}) has the Lorentz decomposition
\begin{equation}
\left\langle\tilde O_{i,\mu\nu\rho\sigma}^\prime
\tilde O_j^{\prime,\mu^\prime\nu^\prime\rho^\prime\sigma^\prime}\right\rangle
=\Pi_{1,ij}^0(q^2)q^2g_{[\mu}^{[\mu^\prime}g_\nu^{\nu^\prime}
g_\rho^{\rho^\prime}g_{\sigma]}^{\sigma^\prime]}
+\Pi_{2,ij}^0(q^2)q_{[\mu}q^{[\mu^\prime}g_\nu^{\nu^\prime}
g_\rho^{\rho^\prime}g_{\sigma]}^{\sigma^\prime]},
\label{cor}
\end{equation}
where $\Pi_{1,ij}^0$ and $\Pi_{2,ij}^0$ are functions of $q^2$.
We may extract $\Pi_{1,ij}^0$ and $\Pi_{2,ij}^0$ by totally contracting
Eq.~(\ref{cor}) with the projectors
\begin{eqnarray}
P_{1,\mu\nu\rho\sigma}^{\mu^\prime\nu^\prime\rho^\prime\sigma^\prime}(q)
&=&\frac{24}{(q^2)^2}\,\frac{\left(
q^2g_{[\mu}^{[\mu^\prime}g_\nu^{\nu^\prime}g_\rho^{\rho^\prime}
g_{\sigma]}^{\sigma^\prime]}
-4q_{[\mu}q^{[\mu^\prime}g_\nu^{\nu^\prime}g_\rho^{\rho^\prime}
g_{\sigma]}^{\sigma^\prime]}\right)}{(D-1)(D-2)(D-3)(D-4)},
\nonumber\\
P_{2,\mu\nu\rho\sigma}^{\mu^\prime\nu^\prime\rho^\prime\sigma^\prime}(q)
&=&\frac{96}{(q^2)^2}\,\frac{\left(
-q^2g_{[\mu}^{[\mu^\prime}g_\nu^{\nu^\prime}g_\rho^{\rho^\prime}
g_{\sigma]}^{\sigma^\prime]}
+Dq_{[\mu}q^{[\mu^\prime}g_\nu^{\nu^\prime}g_\rho^{\rho^\prime}
g_{\sigma]}^{\sigma^\prime]}\right)}{(D-1)(D-2)(D-3)(D-4)},
\label{pro}
\end{eqnarray}
respectively.
Notice that $P_1$ and $P_2$ develop $1/\varepsilon$ poles in the physical
limit $\varepsilon\to0$.
This may be understood by observing that the two terms on the right-hand side 
of Eq.~(\ref{cor}) are actually linearly dependent in four dimensions.
In practice, the appearance of $1/\varepsilon$ poles in Eq.~(\ref{pro}) does
not create a problem, since we are only interested in the absorptive parts
of the correlators in Eq.~(\ref{cor}), so that it suffices to extract the pole 
parts of the relevant diagrams.
It turns out that $\Pi_{1,ij}^0=0$, so that, after performing renormalization,
taking the physical limit $\varepsilon\to0$, and contracting with the
Levi-Civita tensors, we have
\begin{equation}
\left\langle\left[\tilde O_i^\prime\right]\left[\tilde O_j^\prime\right]
\right\rangle=-6q^2\Pi_{2,ij}(q^2),
\end{equation}
where $\Pi_{2,ij}$ is the renormalized version of $\Pi_{2,ij}^0$.
In fact, it can be shown that $\Pi_{1,ij}^0$ vanishes on kinematical grounds.
As is well known, $\tilde O_1^\prime$ can be written as the divergence of the
so-called Chern-Simons current,
$K^{\prime,\mu}=\epsilon^{\mu\nu\rho\sigma}K_{\nu\rho\sigma}^\prime$
with
$K_{\nu\rho\sigma}^\prime=4G_\nu^{0\prime,a}\partial_\rho G_\sigma^{0\prime,a}
+(4/3)g_s^{0\prime}f^{abc}G_\nu^{0\prime,a}G_\rho^{0\prime,b}
G_\sigma^{0\prime,c}$, i.e.\ $\tilde O_1^\prime=\partial_\mu K^{\prime,\mu}$,
which is an exact identity.
This implies that
$\tilde O_{1,\mu\nu\rho\sigma}^\prime=\partial_{[\mu}
K_{\nu\rho\sigma]}^\prime$.
Thus, the correlator in Eq.~(\ref{cor}) is represented by just one term
proportional to
$q_{[\mu}q^{[\mu^\prime}\langle K_{\nu\rho\sigma]}^\prime
K^{\prime,\nu^\prime\rho^\prime\sigma^\prime]}\rangle$,
whence it follows that $\Pi_{1,ij}^0=0$.

\begin{figure}[t]
\leavevmode
\begin{center}
\epsfxsize=16cm
\epsffile[118 552 478 731]{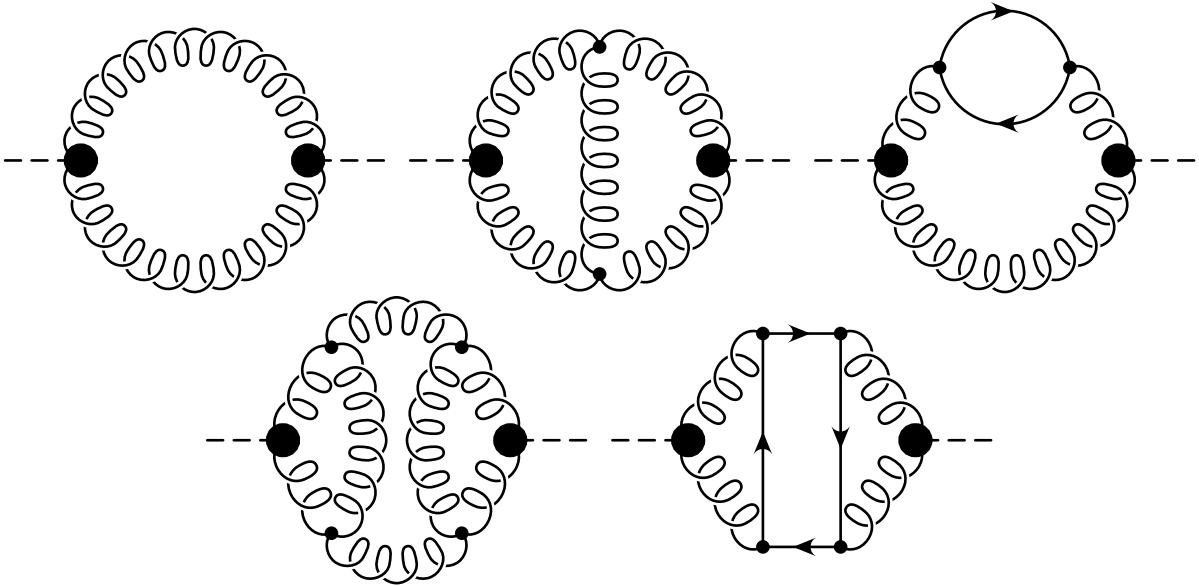}
\caption{Typical Feynman diagrams contributing to the correlator
$\langle\tilde O_1^\prime\tilde O_1^\prime\rangle$.
Looped, solid, and dashed lines represent gluons, light quarks, and $A$ 
bosons, respectively.
Solid circles represent insertions of $\tilde O_1^\prime$.}
\label{fig:two}
\end{center}
\end{figure}

Due to Eq.~(\ref{mix}), all three correlators
$\langle\tilde O_1^\prime\tilde O_1^\prime\rangle$,
$\langle\tilde O_1^\prime\tilde O_2^\prime\rangle$, and
$\langle\tilde O_2^\prime\tilde O_2^\prime\rangle$ contribute to
$\langle[\tilde O_1^\prime][\tilde O_1^\prime]\rangle$, the absorptive part of
which we wish to calculate through ${\cal O}(\alpha_s^2)$.
At the three-loop level, these three correlators receive contributions from
$403$, $28$, and $33$ massless diagrams, respectively.
Typical examples pertaining to
$\langle\tilde O_1^\prime\tilde O_1^\prime\rangle$ are depicted in
Fig.~\ref{fig:two}.
We generate and evaluate the contributing diagrams with the packages QGRAF 
\cite{nog} and MINCER \cite{gor}, which is written in FORM \cite{ver}.
We work in the covariant gauge with arbitrary gauge parameter.
The cancellation of the latter in the final results serves as a welcome check.

Our results for the absorptive parts of the renormalized correlators read
\begin{eqnarray}
{\rm Im}\,
\left\langle\left[\tilde O_1^\prime\right]\left[\tilde O_1^\prime\right]
\right\rangle
&=&\frac{(q^2)^2}{\pi}N_A
\left\{1+\frac{\alpha_s^{(n_l)}(\mu)}{\pi}
\left[C_A\left(\frac{97}{12}+\frac{11}{6}\ln\frac{\mu^2}{q^2}\right)
+Tn_l\left(-\frac{7}{3}
\right.\right.\right.\nonumber\\
&&{}-\left.\left.
\frac{2}{3}\ln\frac{\mu^2}{q^2}\right)\right]
+\left(\frac{\alpha_s^{(n_l)}(\mu)}{\pi}\right)^2
\left[C_A^2\left(\frac{51959}{864}-\frac{121}{24}\zeta(2)-\frac{55}{8}\zeta(3)
\right.\right.\nonumber\\
&&{}+\left.
\frac{1135}{48}\ln\frac{\mu^2}{q^2}+\frac{121}{48}\ln^2\frac{\mu^2}{q^2}
\right)
+C_FTn_l\left(-\frac{107}{12}+3\zeta(3)-2\ln\frac{\mu^2}{q^2}\right)
\nonumber\\
&&{}
+C_ATn_l\left(-\frac{3793}{108}+\frac{11}{3}\zeta(2)-\frac{1}{2}\zeta(3)
-\frac{46}{3}\ln\frac{\mu^2}{q^2}-\frac{11}{6}\ln^2\frac{\mu^2}{q^2}\right)
\nonumber\\
&&{}+\left.\left.
T^2n_l^2\left(\frac{251}{54}-\frac{2}{3}\zeta(2)
+\frac{7}{3}\ln\frac{\mu^2}{q^2}+\frac{1}{3}\ln^2\frac{\mu^2}{q^2}\right)
\right]\right\}
\nonumber\\
&=&
\frac{(q^2)^2}{\pi}8
\left\{1+\frac{\alpha_s^{(n_l)}(\mu)}{\pi}\left[
\frac{97}{4}+\frac{11}{2}\ln\frac{\mu^2}{q^2}
+n_l\left(-\frac{7}{6}-\frac{1}{3}\ln\frac{\mu^2}{q^2}\right)\right]
\right.\nonumber\\
&&{}
+\left(\frac{\alpha_s^{(n_l)}(\mu)}{\pi}\right)^2\left[
\frac{51959}{96}-\frac{363}{8}\zeta(2)-\frac{495}{8}\zeta(3)
+n_l\left(-\frac{469}{8}+\frac{11}{2}\zeta(2)
\right.\right.\nonumber\\
&&{}+\left.
\frac{5}{4}\zeta(3)\right)
+n_l^2\left(\frac{251}{216}-\frac{1}{6}\zeta(2)\right)
+\left(\frac{3405}{16}-\frac{73}{3}n_l+\frac{7}{12}n_l^2\right)
\ln\frac{\mu^2}{q^2}
\nonumber\\
&&{}+\left.\left.
\left(\frac{363}{16}-\frac{11}{4}n_l+\frac{1}{12}n_l^2\right)
\ln^2\frac{\mu^2}{q^2}\right]\right\},
\nonumber\\
{\rm Im}\,
\left\langle\left[\tilde O_1^\prime\right]\left[\tilde O_2^\prime\right]
\right\rangle
&=&\frac{(q^2)^2}{\pi}\,\frac{\alpha_s^{(n_l)}(\mu)}{\pi}
\left(\frac{N_ATn_l}{4}\right),
\nonumber\\
{\rm Im}\,
\left\langle\left[\tilde O_2^\prime\right]\left[\tilde O_2^\prime\right]
\right\rangle
&=&\frac{(q^2)^2}{\pi}\left(\frac{\alpha_s^{(n_l)}(\mu)}{\pi}\right)^2
\left(\frac{N_AT^2n_l^2}{16}\right),
\label{imc}
\end{eqnarray}
where $\zeta$ is Riemann's zeta function, with values $\zeta(2)=\pi^2/6$ and
$\zeta(3)\approx1.202\,057$, and we have put $N_c=3$ in the second expression
on the right-hand side.
Notice that ${\rm Im}\,\langle[\tilde O_2^\prime][\tilde O_2^\prime]\rangle$
starts at ${\cal O}(\alpha_s^2)$.
This may be understood by observing that the $A\to q_i\bar q_i$ decay width is
helicity suppressed and quenched if the quark $q_i$ is taken to be massless.
Thus, in order for a diagram to contribute it must have a cut which only
involves gluons.
Such diagrams first appear at three loops.
Actually, the results in Eq.~(\ref{imc}) are not mutually independent.
In fact, Eq.~(\ref{ano}) allows us to derive
${\rm Im}\,\langle[\tilde O_1^\prime][\tilde O_2^\prime]\rangle$ and
${\rm Im}\,\langle[\tilde O_2^\prime][\tilde O_2^\prime]\rangle$ from
${\rm Im}\,\langle[\tilde O_1^\prime][\tilde O_1^\prime]\rangle$ and thus
provides a welcome check for Eq.~(\ref{imc}).
The ${\cal O}(\alpha_s)$ term of
${\rm Im}\,\langle[\tilde O_1^\prime][\tilde O_1^\prime]\rangle$ in
Eq.~(\ref{imc}) is in agreement with Ref.~\cite{kat}.

All ingredients which enter the calculation of the $A\to gg$ decay width 
through ${\cal O}(\alpha_s^4)$ are now available.
From Eq.~(\ref{ren}), we derive the general expression
\begin{equation}
\Gamma(A\to gg)=\frac{\sqrt2G_F}{M_A}\left(
\tilde C_1^2{\rm Im}\,\left\langle\left[\tilde O_1^\prime\right]
\left[\tilde O_1^\prime\right]\right\rangle
+2\tilde C_1\tilde C_2{\rm Im}\,\left\langle\left[\tilde O_1^\prime\right]
\left[\tilde O_2^\prime\right]\right\rangle
+\tilde C_2^2{\rm Im}\,\left\langle\left[\tilde O_2^\prime\right]
\left[\tilde O_2^\prime\right]\right\rangle\right),
\label{gam}
\end{equation}
where it is understood that the correlators are to be evaluated at
$q^2=M_A^2$.
The last term contained within the parenthesis of Eq.~(\ref{gam}) contributes
in ${\cal O}(\alpha_s^6)$ and is only included for completeness.
Inserting Eqs.~(\ref{con}) and (\ref{imc}) in Eq.~(\ref{gam}) and putting
$N_c=3$, $n_l=5$, and $\mu=M_A$, we finally obtain
\begin{eqnarray}
\Gamma(A\to gg)&=&\frac{G_FM_A^3}{16\pi\sqrt2}
\left(\frac{\alpha_s^{(5)}(M_A)}{\pi}\right)^2
\left[1+\frac{221}{12}\,\frac{\alpha_s^{(5)}(M_A)}{\pi}
+\left(\frac{\alpha_s^{(5)}(M_A)}{\pi}\right)^2
\right.\nonumber\\
&&{}\times\left.
\left(\frac{237311}{864}-\frac{529}{24}\zeta(2)-\frac{445}{8}\zeta(3)
-5\ln\frac{M_t^2}{M_A^2}\right)\right]
\nonumber\\
&\approx&\frac{G_FM_A^3}{16\pi\sqrt2}
\left(\frac{\alpha_s^{(5)}(M_A)}{\pi}\right)^2
\left[1+18.417\,\frac{\alpha_s^{(5)}(M_A)}{\pi}
+\left(\frac{\alpha_s^{(5)}(M_A)}{\pi}\right)^2
\right.\nonumber\\
&&{}\times\left.
\left(171.544-5\ln\frac{M_t^2}{M_A^2}\right)\right].
\label{num}
\end{eqnarray}
The ${\cal O}(\alpha_s)$ correction in Eq.~(\ref{num}) agrees with the result
originally found in Refs.~\cite{spi,kat}.
If we assume that $\alpha_s^{(5)}(M_A)=0.116$, which follows from
$\alpha_s^{(5)}(M_Z)=0.118$ for $M_A=100$~GeV, and take the $t$-quark pole 
mass to be $M_t=175.6$~GeV, then the correction factor corresponding to the
square bracket in Eq.~(\ref{num}) has the value $1+0.68+0.23=1.91$,
i.e.\ the three-loop term amounts to 33\% of the two-loop term.
This is somewhat larger than the corresponding correction factor for a SM
Higgs boson with mass $M_H=100$~GeV, which was found to be $1+0.66+0.21=1.87$
\cite{che}.
For our choice of input parameters, we obtain from Eq.~(\ref{num}) the
QCD-corrected prediction $\Gamma(A\to gg)=426$~keV.

Similarly to the $H\to gg$ case \cite{che}, Eq.~(\ref{num}) may be RG improved
by resumming the logarithms of the type $\ln(M_t^2/M_A^2)$.
This leads to
\begin{eqnarray}
\Gamma(A\to gg)&=&\frac{G_FM_A^3}{16\pi\sqrt2}
\left(\frac{\alpha_s^{(5)}(M_A)}{\pi}\right)^2
\left[1+\frac{4363}{276}\,\frac{\alpha_s^{(5)}(M_A)}{\pi}
+\frac{60}{23}\,\frac{\alpha_s^{(6)}(M_t)}{\pi}
\right.\nonumber\\
&&{}+\left(\frac{239471}{864}-\frac{529}{24}\zeta(2)-\frac{445}{8}\zeta(3)
\right)\left(\frac{\alpha_s^{(5)}(M_A)}{\pi}\right)^2
\nonumber\\
&&{}-\left.
\frac{1800}{529}\,\frac{\alpha_s^{(5)}(M_A)}{\pi}\,
\frac{\alpha_s^{(6)}(M_t)}{\pi}
+\frac{955}{1058}\,\left(\frac{\alpha_s^{(6)}(M_t)}{\pi}\right)^2\right]
\nonumber\\
&\approx&\frac{G_FM_A^3}{16\pi\sqrt2}
\left(\frac{\alpha_s^{(5)}(M_A)}{\pi}\right)^2
\left[1+15.808\,\frac{\alpha_s^{(5)}(M_A)}{\pi}
+2.609\,\frac{\alpha_s^{(6)}(M_t)}{\pi}
\right.\nonumber\\
&&{}+174.044\left(\frac{\alpha_s^{(5)}(M_A)}{\pi}\right)^2
-3.403\,\frac{\alpha_s^{(5)}(M_A)}{\pi}\,\frac{\alpha_s^{(6)}(M_t)}{\pi}
\nonumber\\
&&{}+\left.
0.903\left(\frac{\alpha_s^{(6)}(M_t)}{\pi}\right)^2\right].
\end{eqnarray}
Since, at present, the experimental lower bound on $M_A$ in the MSSM is 
24.3~GeV \cite{pdg}, we have $\ln(M_t^2/M_A^2)<4$, so that the numerical
effect of the RG improvement is negligible in practical applications.

It is interesting to compare the exact value of the ${\cal O}(\alpha_s^2)$
correction in Eq.~(\ref{num}) with the estimates one may derive from the
knowledge of the ${\cal O}(\alpha_s)$ correction through the application of
well-known scale-optimization procedures, based on the fastest apparent
convergence (FAC) \cite{fac}, the principle of minimal sensitivity (PMS)
\cite{pms}, and the proposal by Brodsky, Lepage, and Mackenzie (BLM)
\cite{blm} to resum the leading light-quark contribution to the
renormalization of the strong coupling constant.
These procedures lead to the generic expression
\begin{eqnarray}
\Gamma(A\to gg)&=&\frac{G_FM_A^3}{16\pi\sqrt2}
\left(\frac{\alpha_s^{(5)}(\xi M_A)}{\pi}\right)^2
\left(1+\bar K_1\frac{\alpha_s^{(5)}(\xi M_A)}{\pi}\right)
\nonumber\\
&=&\frac{G_FM_A^3}{16\pi\sqrt2}
\left(\frac{\alpha_s^{(5)}(M_A)}{\pi}\right)^2
\left[1+K_1\frac{\alpha_s^{(5)}(M_A)}{\pi}
+\bar K_2\left(\frac{\alpha_s^{(5)}(M_A)}{\pi}\right)^2\right],
\label{opt}
\end{eqnarray}
where $K_1=k_1+n_l\kappa_1$, with $k_1=97/4$ and $\kappa_1=-7/6$, is the
coefficient of the ${\cal O}(\alpha_s)$ correction in Eq.~(\ref{num}).
The FAC, PMS, and BLM expressions for $\xi$, $\bar K_1$, and $\bar K_2$ read
\begin{eqnarray}
\ln\xi^{\rm FAC}&=&-\frac{K_1}{4\beta_0},\qquad
\bar K_1^{\rm FAC}=0,\qquad
\bar K_2^{\rm FAC}=K_1\left(\frac{3}{4}K_1+\frac{\beta_1}{\beta_0}\right),
\nonumber\\
\ln\xi^{\rm PMS}&=&
-\frac{1}{2\beta_0}\left(\frac{K_1}{2}+\frac{\beta_1}{3\beta_0}\right),\qquad
\bar K_1^{\rm PMS}=-\frac{2\beta_1}{3\beta_0},\qquad
\bar K_2^{\rm PMS}
=\frac{1}{3}\left(\frac{3}{2}K_1+\frac{\beta_1}{\beta_0}\right)^2,
\nonumber\\
\ln\xi^{\rm BLM}&=&\frac{3}{2}\kappa_1,\qquad
\bar K_1^{\rm BLM}=k_1+\frac{33}{2}\kappa_1,
\nonumber\\
\bar K_2^{\rm BLM}&=&\kappa_1\left[\frac{3}{4}\kappa_1n_l^2
+\frac{1}{2}\left(3k_1+\frac{19}{2}\right)n_l
-\frac{9}{4}\left(11k_1+\frac{363}{4}\kappa_1+17\right)\right],
\label{val}
\end{eqnarray}
respectively, where $\beta_0=11/4-n_l/6$ and $\beta_1=51/8-19n_l/24$ are the
first two coefficients of the Callan-Symanzik beta function of QCD.
The numerical results for $n_l=5$ are summarized in Table~\ref{tab:one}.
The values of $\bar K_2$ should be compared with the true coefficient $K_2$ of
the ${\cal O}(\alpha_s^2)$ correction in Eq.~(\ref{num}).
For completeness, we also list in Table~\ref{tab:one} the corresponding
results for the $H\to gg$ decay width.
In this case, one has $K_2\approx156.808-5.708\ln(M_t^2/M_H^2)$ \cite{che}.
In both cases, all three scale-optimization prescriptions correctly predict 
the sign and the order of magnitude of $K_2$.
Furthermore, the three $\bar K_2$ values for the $H\to gg$ decay width are
indeed smaller than the respective values for the $A\to gg$ case.
Similarly to Ref.~\cite{sir}, the FAC and PMS results almost coincide.

\begin{table}[t]
\begin{center}
\caption{Numerical evaluation of Eq.~(\ref{val}) with $n_l=5$ for the
$A\to gg$ and $H\to gg$ decay widths.}
\label{tab:one}
\medskip
\begin{tabular}{|c|c|c|c|c|c|c|} \hline\hline
& \multicolumn{3}{c|}{$A\to gg$} & \multicolumn{3}{c|}{$H\to gg$} \\
\cline{2-7}
\rule{0mm}{5mm}&
$\xi$ & $\bar K_1$ & $\bar K_2$ & $\xi$ & $\bar K_1$ & $\bar K_2$ \\
\hline
FAC & 0.091 & 0 & 277.601 & 0.097 & 0 & 263.346 \\
PMS & 0.081 & $-0.841$ & 278.131 & 0.087 & $-0.841$ & 263.876 \\
BLM & 0.174 & 5 & 252.547 & 0.174 & 4.5 & 242.484 \\
\hline\hline
\end{tabular}
\end{center}
\end{table}

\section{\label{sec:four}Discussion and conclusions}

In this paper, we studied the interactions of a neutral CP-odd scalar boson
$A$ with gluons and $n_l$ light quarks $q_i$ in the presence of a heavy quark
$t$, with mass $m_t\gg M_A/2$, through three loops in QCD. 
For simplicity, we assumed that the Yukawa couplings of the light quarks may
be neglected, which is a useful approximation in the 2HDM with
$\tan\beta\approx1$, where the Yukawa couplings and the masses of the quarks
are related by a flavour-independent proportionality factor.
Starting from the Yukawa Lagrangian~(\ref{ful}) embedded in full QCD, we
integrated out the $t$ quark to obtain the corresponding effective Lagrangian
of the $n_l$-flavour theory, the renormalized version of which is given by 
Eq.~(\ref{ren}).
The operators in Eq.~(\ref{ren}) comprise only light fields, while all 
residual dependence on the $t$ quark is contained in their coefficient 
functions.
We diagrammatically evaluated the coefficients $\tilde C_1$ and $\tilde C_2$
of the physical operators $[\tilde O_1^\prime]$ and $[\tilde O_2^\prime]$
through ${\cal O}(\alpha_s^3)$ and ${\cal O}(\alpha_s^2)$, respectively.
This is consistent, since $[\tilde O_2^\prime]$ is of ${\cal O}(\alpha_s)$
relative to $[\tilde O_1^\prime]$ as may be seen from Eq.~(\ref{ano}).
We worked in the $\overline{\rm MS}$ renormalization scheme \cite{bur} with
the convention that Eq.~(\ref{ano}) should be exact \cite{lar}.
Our results for $\tilde C_1$ and $\tilde C_2$ are given in Eqs.~(\ref{con})
and (\ref{ctw}), respectively.
In particular, we found that the ${\cal O}(\alpha_s)$ and
${\cal O}(\alpha_s^2)$ corrections to $\tilde C_1$ exactly vanish if the
latter is expressed in terms of $\alpha_s^{(n_l)}(\mu)$.
We thus verified, by explicit calculation through three loops, the all-order 
prediction that $\tilde C_1$ does not receive any QCD corrections, which
follows via a low-energy theorem \cite{spi,djo} from the nonabelian version
\cite{wab} of the Adler-Bardeen non-renormalization theorem \cite{adl}.

At this point, we should emphasize that, due to Eq.~(\ref{ano}), the
distinction between the renormalized operators $[\tilde O_1^\prime]$ and
$[\tilde O_2^\prime]$ does not have a deep physical meaning.
Actually, from Eqs.~(\ref{ren}) and (\ref{ano}) it follows that the physical
coupling of the $A$ boson to $[\tilde O_1^\prime]$ is proportional to
$[\tilde C_1+\alpha_s^{(n_l)}(\mu)Tn_l\tilde C_2/(4\pi)]$.
Thus, the low-energy theorem gets violated starting from
${\cal O}(\alpha_s^3)$ unless we stick to the $\overline{\rm MS}$ definitions
of $[\tilde O_1^\prime]$ and $[\tilde O_2^\prime]$ \cite{lar}.

The effective Lagrangian~(\ref{ren}) allows us to evaluate physical
observables related to the interactions of the $A$ boson with gluons and light
quarks to higher orders in QCD by just considering massless diagrams in the
$n_l$-flavour theory.
As an application, we evaluated the ${\cal O}(\alpha_s^2)$ correction to the
$A\to gg$ decay width, extending the result of Refs.~\cite{spi,kat} by one 
order.
Our final result is listed in Eq.~(\ref{num}).
For $M_A=100$~GeV, the overall QCD correction factor turned out to be as large 
as $1+0.68+0.23=1.91$.
It is slightly larger than the corresponding correction factor for the
$H\to gg$ decay width of the SM Higgs boson $H$ with the same mass, which was 
found to be $1+0.66+0.21=1.87$ \cite{che}.
The FAC \cite{fac}, PMS \cite{pms}, and BLM \cite{blm} predictions for the
${\cal O}(\alpha_s^2)$ correction to the $A\to gg$ decay width are
0.38, 0.38, and 0.34, respectively.
The corresponding results for the $H\to gg$ case are 0.36, 0.36, and 0.33.

%\vspace{1cm}
\newpage
\noindent
{\bf Acknowledgements}
\smallskip

\noindent
We thank S.J. Brodsky and P.A. Grassi for useful discussions.
W.A.B. and K.G.C. thank the MPI Theory Group for the hospitality extended to
them during visits when the research reported here was initiated.
B.A.K. thanks the NYU Physics Department for the hospitality extended to him 
during a visit when this manuscript was finalized.
The work of K.G.C. was supported in part by INTAS under Contract
INTAS--93--744--ext and by DFG under Contract KU~502/8--1.

%\newpage

\end{document}